\documentclass[twocolumn]{aastex62}
\usepackage{amssymb}
\usepackage{amsmath}
\usepackage{amssymb}
\usepackage{empheq}
\usepackage{mathrsfs}
\usepackage{wasysym} 
\usepackage{etoolbox}
\usepackage{lmodern}

\newcommand{\be}{\begin{equation}}
\newcommand{\ee}{\end{equation}}
\newcommand{\G}{\mathcal{G}} 
 
\newcommand{\B}{B_{\star}}

\begin{document} 

\title{\bf From Tides to Currents:\\Unraveling the Mechanism That Powers WASP-107b's Internal Heat Flux}

\author[0000-0002-7094-7908]{Konstantin Batygin}
\affiliation{Division of Geological and Planetary Sciences, California Institute of Technology, Pasadena, CA 91125}

\begin{abstract}
The sub-Jovian exoplanet WASP-107b ranks among the best-characterized low-density worlds, featuring a Jupiter-like radius and a mass that lies firmly in the sub-Saturn range. Recently obtained JWST spectra reveal significant methane depletion in the atmosphere, indicating that WASP-107b’s envelope has both a high metallicity and an elevated internal heat flux. Together with a detected non-zero orbital eccentricity, these data have been interpreted as evidence of tidal heating. However, explaining the observed luminosity with tidal dissipation requires an unusually low tidal quality factor of $Q \sim 100$. Moreover, we find that secular excitation by the RV-detected outer companion WASP-107c, generally cannot sustain WASP-107b’s eccentricity in steady state against tidal circularization. As an alternative explanation, we propose that Ohmic dissipation — generated by interactions between zonal flows and the planetary magnetic field in a partially ionized atmosphere — maintains the observed thermal state. Under nominal assumptions for the field strength, atmospheric circulation, and ionization chemistry, we show that Ohmic heating readily accounts for WASP-107b’s inflated radius and anomalously large internal entropy. 
\end{abstract}

\keywords{Exoplanets, Planetary dynamics, Planetary structure}

\section{Introduction}
\label{sec:intro} 



Over the course of the past three decades, the discovery and characterization of extrasolar planets have unveiled a steady succession of unexpected phenomena, challenging and advancing our understanding of planetary formation, structure, and evolution \citep{1992Natur.355..145W,1995Natur.378..355M}. Among the earliest surprises was the larger-than-expected radii of close-in giant planets, first exemplified by the detection of HD 209458b in transit \citep{2000ApJ...529L..45C,2000ApJ...529L..41H}. subsequent observations revealed that radii exceeding that of Jupiter are a common feature of short-period giant planets, and the question of how these objects maintain anomalously high interior entropy despite rapid radiative heat loss came to be recognized as the hot Jupiter inflation problem.

To explain this inflation, several energy injection mechanisms have been proposed, including tidal dissipation \citep{2001ApJ...548..466B}, kinetic heating \citep{2002A&A...385..156G}, thermal tides \citep{2010ApJ...714....1A}, turbulent burial of atmospheric heat \citep{2010ApJ...721.1113Y}, Ohmic dissipation \citep{BS10}, and vertical advection of potential temperature \citep{2017ApJ...841...30T}. In parallel, mechanisms that slow gravitational contraction, such as enhanced atmospheric metallicity \citep{2007ApJ...661..502B} have also been suggested. While each of these likely contributes to some extent (e.g., \citealt{2021A&A...645A..79S}), statistical studies of the exoplanet population point towards Ohmic dissipation as the dominant process driving hot Jupiter inflation \citep{2011ApJ...729L...7L,2018AJ....155..214T}.

As the demographics of lower-mass exoplanets came into sharper focus, it became apparent that sub-Jovian planets can also display anomalously large (i.e., $R \sim R_{\jupiter}$) radii. Standard planetary structure calculations indicate that old (ages $\gtrsim1\,$Gyr), intermediate-mass ($\sim10-50\,M_{\oplus}$) planets can achieve Jupiter-like radii only if they are predominantly composed of hydrogen and helium, with minimal heavy-element cores \citep{2013ApJ...769L...9B,2014ApJ...792....1L}, or if they are subject to ancillary interior heating. Both explanations, however, require conditions that extend beyond standard models of planetary formation and interior evolution: nearly pure H/He compositions pose a substantial problem within the context of the core accretion theory, which fundamentally requires the presence of massive $(\gtrsim10\,M_{\oplus})$ cores to initiate meaningful gas accretion during the limited lifetimes of protoplanetary disks \citep{1996Icar..124...62P}. On the other hand, internal heating scenarios necessitate sustained mechanisms capable of continuously depositing energy into the convective interior of the planet. Consequently, the large radii of inflated sub-Jovian planets present an intriguing problem, as they must either result from substantially reduced bulk metallicities -- contradicting the predictions of core-accretion -- or from persistent internal heating, which would manifest observationally as an elevated intrinsic heat flux.

A notable example that illustrates this tension is WASP-107b, a ``super-puff" planet with a mass of $m_b = 30.5\pm1.7\,M_{\oplus}$ and a radius comparable to Jupiter ($R_b = 0.95\pm0.03\,R_{\jupiter}$), orbiting a $M_{\star} = 0.683 \pm 0.002\,\,M_{\odot}$ star, on a $P_b=5.7$ day orbit\footnote{The corresponding equilibrium temperature of the planet is $T_{\rm{irr}}=770$\,K)}. While this large radius was initially attributed to a near-unity H/He envelope mass fraction (see e.g.,  \citealt{2021AJ....161...70P} and references therein), more recently, two independent studies \citep{Sing,Welbanks} analyzed JWST spectra of WASP-107b, and reported a significant depletion of CH$_4$ relative to equilibrium expectations, indicating a high intrinsic heat flux (thus signaling inflation). Mechanistically, this inference is enabled by the fact that the abundance of methane in planetary atmospheres is partially governed by the chemical reaction $\mathrm{CH_4 + H_2O \rightleftharpoons CO + 3\,H_2}$, in which the right-hand side is favored at high temperatures. Therefore, the observed under-abundance of $\mathrm{CH_{4}}$ relative to $\mathrm{CO}$ points to a hotter-than-expected atmosphere that is heated from below.

Employing detailed models, the retrievals of \citet{Sing} yielded an effective interior temperature of $T_{\rm{eff}} = 460\pm40\,$K for WASP-107b, a heavy-element core mass of $M_{\rm{c}} = 11.5^{+3}_{-3.5} \, M_{\oplus}$, and an atmospheric metallicity of $43\pm8$ times the solar value (equivalent to an envelope heavy-element fraction $Z_{\rm{env}} \approx 0.37^{+0.05}_{-0.04}$, and a total heavy-element mass of $\sim 19 M_{\oplus}$). \citet{Welbanks} found consistent results, estimating $T_{\rm{eff}} > 345\,K$ and a total heavy-element mass of $\sim 22\,M_{\oplus}$, which they assumed to be largely concentrated in a solid core.

Beyond structural considerations, both of the aforementioned studies attribute WASP-107b’s high intrinsic heat flux to tidal heating. Tidal dissipation has previously been shown to explain inflated envelopes in several sub-Saturn-class planets, such as those in the K2-19 system \citep{2020AJ....159....2P,Mi2020ApJ...897....7M}, and radial velocity measurements of WASP-107b indeed suggest a significant eccentricity ($e_b = 0.06\pm0.04$; \citealt{2021AJ....161...70P}) that could in principle drive the necessary tidal luminosity. Additionally, the WASP-107 system contains a mildly eccentric ($e_c = 0.28\pm0.07$), long-period ($P_c = 2.98\pm0.04\,$yr) giant-planet companion with a minimum mass of $m_c \sin(i_{\rm{sky}}) = 0.36\pm0.04 M_{\jupiter}$, potentially allowing for the sustained eccentricity of WASP-107b through secular perturbations. This hypothesis is further strengthened by a Rossiter-McLaughlin measurement indicating WASP-107b’s near-polar projected orbital obliquity of $\psi = 118^{+38}_{-19}\,\deg$ \citep{2017AJ....153..205D,2021AJ....161..119R}, thus hinting at a bonafide three-dimensional orbital configuration\footnote{\citet{2024ApJ...972..159Y} have examined a high-eccentricity migration scenario for the origin of WASP-107b, and have predicted a mutual inclination between the planetary orbits in excess of $25\,\deg$.} that would allow planet c’s true mass to significantly exceed its minimum estimate.

A primary goal of this work is to assess the tidal heating hypothesis more rigorously and to consider an alternative mechanism. In section \ref{sec:tidal}, we employ a series of analytic and numeric calculations to demonstrate that, despite their attractiveness, eccentricity tides are unlikely to explain WASP-107b’s properties, given the current observational constraints. Instead, in section \ref{sec:ohmic}, we propose Ohmic dissipation as the principal driver of the planet’s intrinsic luminosity, and support this argument with a series of semi-analytic estimates. Finally, in section \ref{sec:discussion}, we summarize our findings and discuss their broader implications for highly inflated sub-Jovian exoplanets.

\section{Tidal Heating}
\label{sec:tidal} 


\subsection{Preliminaries}
We begin our analysis of the tidal heating hypothesis by establishing some preliminaries. Given a planetary specific dissipation function $Q_b$ and assuming spin-orbit pseudo-synchronization, the interior tidal luminosity can be expressed as:
\begin{align}
L_b = \frac{21}{2}\frac{k_{2b}}{Q_b}\frac{\G\,M_{\star}^2}{a_b} \bigg( \frac{R_b}{a_b} \bigg)^5\,n_b\,e_b^2,
\label{Ltidal}
\end{align}
where $n_b$ denotes the mean motion and $k_{2b}$ is the planetary Love number, representing the degree of central concentration for fluid planets. Relating this luminosity to the observed interior heat flux, $L_b=4\,\pi\,R_b^2\,\sigma_{\rm{sb}}\,T_{\rm{eff}}^4$ and adopting best-fit atmospheric and orbital parameters of \citet{Sing,2021AJ....161...70P}, we find $k_{2b} / Q_b = 5.3 \times 10^{-4}$. 

\begin{figure*}[t]
\centering
\includegraphics[width=\textwidth]{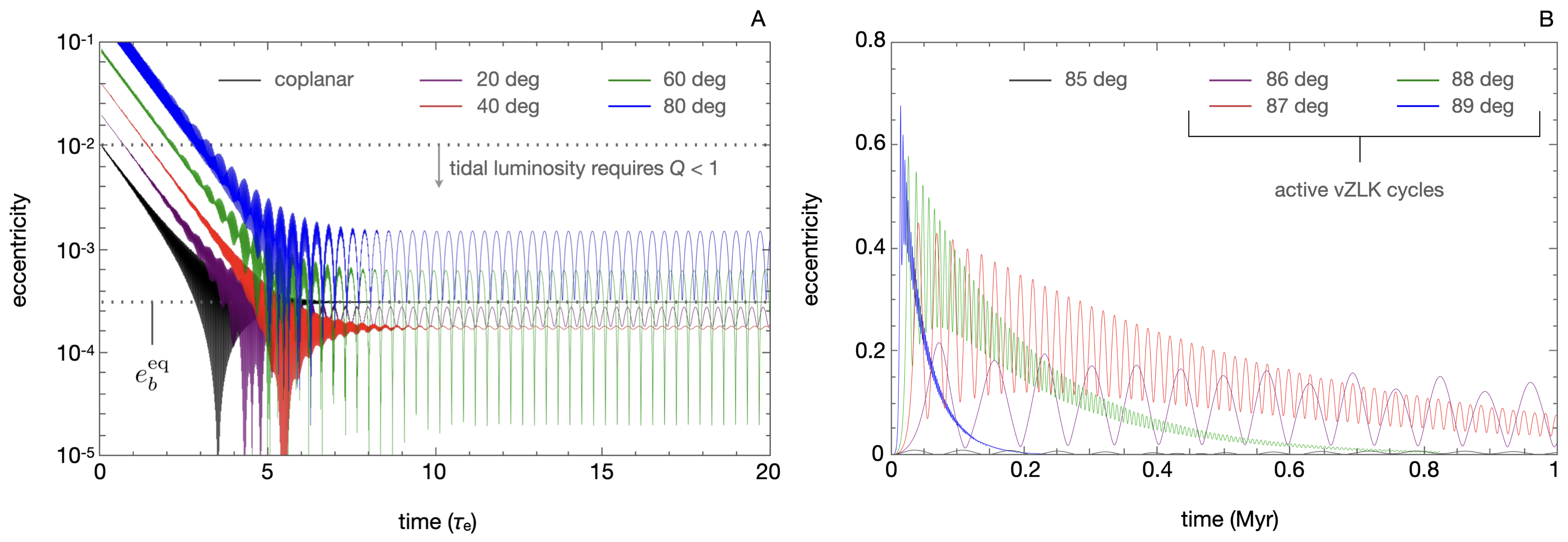}
\caption{Eccentricity evolution of WASP-107b under various dynamical configurations. Panel A depicts the results from secular perturbation theory, showing the orbital eccentricity of WASP-107b as a function of time (in units of the tidal circularization timescale $\tau_e\approx6$\,Myr) for different mutual inclinations of the planetary orbits: $0\deg$ (black), $20\deg$ (purple), $40\deg$ (red), $60\deg$ (green), and $80\deg$ (blue). The coplanar case settles into an equilibrium eccentricity $e_b^{\rm eq} \approx 3 \times 10^{-4}$, well below the threshold ($e_b \gg 0.006$) needed to sustain the observed internal luminosity via tidal dissipation. Inclined configurations yield modest limit-cycle $e_b$ fluctuations, but still fail to reach sufficiently large eccentricities. Panel B shows the results of direct N-body simulations for nearly orthogonal configurations with $\Delta\,i=85\deg$ (black), $86\deg$ (purple), $87\deg$ (red), $88\deg$ (green), and $89\deg$ (blue), spanning $1\,$Myr. Although extreme mutual inclinations can trigger large, short-lived eccentricity oscillations through vZLK cycles, these configurations rapidly diverge from the observed parameters: tidal friction reduces the semi-major axis and causes the mutual inclination to decay, thereby quenching the vZLK mechanism. Even fine-tuned, highly inclined scenarios cannot maintain large eccentricities in a self-consistent manner over Gyr timescales. Taken together, these results suggest that tidal heating alone is insufficient to explain WASP-107b’s inferred heat-flux and present-day radius.}
\label{fig:ecc}
\end{figure*}

To refine this relationship, we calculate $k_{2b}$ by modeling the planet’s interior as a $12\,M_{\oplus}$ solid core surrounded by a polytropic envelope (Appendix A). Using a polytropic index of $\zeta \approx  7/2$, we find $k_{2b} = 0.015$, implying $Q_b \approx 30$. It is worth noting that this result is relatively insensitive to parameter variations: for example, assuming a $22\,M_{\oplus}$ solid core\footnote{In this case, under the assumption of a nearly solar-composition envelope, matching the total planetary mass requires a slightly modified polytropic index of $\zeta = 15/4$.} while reducing the intrinsic temperature to $T_{\rm{eff}}=345\,$K (the lower bound obtained by \citealt{Welbanks}), yields $Q_b =34$. Moreover, because radial velocity measurements constrain the eccentricity of WASP-107b to $e \lesssim 0.1$, the upper bound on $Q_b$ is only a factor of $\sim3$ higher than the nominal value, and remains orders of magnitude lower than values inferred for Solar System giants\footnote{An important exception to this rule may be Saturn, where resonance-locking can yield an effective dissipation function of $Q\sim100$ (\citealt{2016MNRAS.458.3867F,2020NatAs...4.1053L}; see also \citealt{2024Icar..41316014G}).} \citep{2009Natur.459..957L,1988Icar...74..172T}.

Although this bound on $Q_b$ is not physically unfeasible\footnote{Recall that the Earth's tidal quality factor is $Q_{\oplus}\approx12$.}, it is essential to keep in mind that tidal heating is sustained only as long as the planet remains eccentric. A direct consequence of tidal energy dissipation is orbital circularization, which occurs on a timescale
\begin{align}
\tau_{e}= \frac{P_b}{21\,\pi}\frac{Q_b}{k_{2b}}\frac{m_b}{M_{\star}} \bigg(\frac{a_b}{R_b} \bigg)^5.
\label{taue}
\end{align}
With $Q_b/k_{2b}$ constrained by $T_{\rm{eff}}$, time timescale evaluates to $\tau_{e} \lesssim 2\,$Myr. Given that $\tau_e$ is considerably shorter than the system's age (inferred to be $3.4\pm0.3\,$Gyr by \citealt{2021AJ....161...70P}), the tidal heating hypothesis would require continual perturbations from WASP-107c to maintain a non-zero eccentricity for WASP-107b in face of circularization. Moreover, because $Q_b$ must exceed unity by a substantial margin, equation (\ref{Ltidal}) demands that this gravitational coupling sustains WASP-107b’s eccentricity well above $e_b \gg 0.01$. Let us now examine this possibility in detail.

\subsection{Tidally-Relaxed Equilibrium and Limit-Cycle Dynamics}

Owing to a large orbital separation between planets b and c ($\alpha = a_b/a_c \ll 1$), and their moderate values of orbital eccentricities, long-term gravitational interactions between planets b and c unfold in the secular regime. Written in terms of scaled Poincar\'e action-angle variables ($\Gamma \approx e_b^2/2, \gamma = -\varpi_b, Z= 1 - \cos(i_b), z=-\Omega_b$, where $\varpi_b$ and $\Omega_b$ are the longitude of periastron and ascending node of planet b respectively, while $i_b$ is its inclination with respect to the orbit of planet c, which is taken to coincide with the reference plane), the Hamiltonian governing the phase-averaged evolution of planet b is (Appendix B):
\begin{align}
&\mathcal{H} = -\frac{n_b}{4}\frac{m_c}{M_{\star}}\frac{\alpha^3}{\epsilon_c^3}\Bigg(\frac{\big(1+3\,\Gamma\big)\,\big(3\,(1-Z)^2-1\big)}{2} \nonumber \\
&+\frac{15\,\Gamma\,\big(1-(1-Z)^2 \big)}{2}\,\cos\big(2(\gamma-z) \big) -\frac{15}{4}\frac{\alpha\,e_c\,\sqrt{2\,\Gamma}}{\epsilon_c^2} \nonumber \\
&\times\Bigg(\frac{\big(2-Z\big)\,\big( 15\, Z^2-20\,Z+4\big)}{8}\,\cos\big(\gamma + \varpi_c\big) \nonumber \\
&+\frac{Z\,\big(15\,Z^2-40\,Z+24 \big)}{8}\,\cos\big(2\,z-\gamma+\varpi_c \big)\Bigg)\Bigg) \nonumber \\
&-\bigg(3\,n_b\frac{\G\,M_{\star}}{a_b\,c^2} + \frac{15}{2}\,n_b\,k_{2b}\,\bigg( \frac{R_b}{a_b} \bigg)^5 \frac{M_{\star}}{m_b} \bigg)\,\Gamma,
\label{Hamiltonian}
\end{align}
where $\epsilon_c = \sqrt{1-e_c^2}$, $c$ is the speed of light, and
\begin{align}
\varpi_c \approx \frac{3}{8}\,\frac{m_b}{M_{\star}}\frac{\alpha^2}{\epsilon_c^4}\big(3\,\cos(i_b)^2-1 \big)\,n_c\,t
\label{varpic}
\end{align}
is the periastron of the outer orbit.

Qualitatively, the terms on the first and second lines of the above expression govern free precession and the von Zeipel-Lidov-Kozai mechanism, respectively. The third and fourth lines denote octupole-level harmonics that govern Runge-Lenz vector coupling and mixed eccentricity-inclination dynamics, while the final term facilitates precession due to general relativity and the quadrupole of the planetary tidal bulge\footnote{Between these two effects, relativistic apsidal advance dominates by about an order of magnitude. Moreover, given the observationally inferred system parameters, planetary rotational and stellar quadrupolar corrections to the precession are negligible (e.g., \citealt{2009ApJ...698.1778R,2009ApJ...704L..49B}), so we disregard them for simplicity.}. The long-term orbital evolution of planet b can be obtained by numerically solving Hamilton’s equations, while accounting for tidal eccentricity damping with a trivial augmentation: $\dot{\Gamma}=-\partial\,\mathcal{H}/\partial\,\gamma-2\,\Gamma/\tau_e$.

\medskip

Let us consider the simple case of co-planar orbits first. In this configuration, it can be shown that over extended timescales ($t \gtrsim 3 \tau_e$), the planetary apsidal lines lock into co-precession and the orbital eccentricity settles into an equilibrium\footnote{This equilibrium corresponds to the fixed point of the Hamiltonian, and is obtained by setting the apsidal alignment condition $\varpi_c=-\gamma$ and then solving the resulting equation $\partial \mathcal{H}/ \partial \Gamma = - \dot{\varpi_c}$ for $\Gamma$.} given by \citep{2007MNRAS.382.1768M}:
\begin{align}
e_b^{\rm{eq}}=\frac{5}{4}\frac{\alpha\,e_c\,\epsilon_c^{-2}}{\big|1-\sqrt{\alpha}\,(m_b/m_c)\,\epsilon_c^{-1} +\upsilon\,\epsilon_c^3 \big|},
\label{ebeq}
\end{align}
where $\upsilon = 4\,(n_b\,a_b/c)^2\,(M_{\star}/m_c)\,(1/\alpha)^3  +10\,k_{2b}\,(R_b/a_b)^5 \allowbreak \times (1/\alpha)^3\,M_{\star}^2/(m_b\,m_c)$ is a measure of the relativistic and quadrupolar contribution to the apsidal precession of the inner orbit, relative to that forced by perturbations from planet c. For nominal parameters, the equilibrium eccentricity evaluates to $e_b^{\rm{eq}} \approx 3 \times 10^{-4}$ -- more than an order of magnitude smaller than the lower bound insinuated by equation (\ref{Ltidal}). Clearly, a coplanar configuration cannot plausibly accommodate the requisite tidal luminosity. 

For an inclined perturber, the equilibrium breaks out into a limit-cycle in the $(e,\Delta \varpi)$ plane, whose frequency approximately corresponds to the circulation of the von Zeipel-Lidov-Kozai harmonic. To evaluate the amplitude of this limit cycle for the WASP-107 system, we integrated the equations of motion derived from Hamiltonian (\ref{Hamiltonian}), for mutual inclinations of $ i_b = 20, 40, 60$, and $80 \,\deg$, increasing the mass of planet c accordingly. The resulting eccentricity time-series are plotted in Figure \ref{fig:ecc}A over $20\,\tau_e$. As shown, none of these solutions attain the necessary eccentricity.


\subsection{von Zeiplel-Lidov-Kozai Dynamics}

It is of course possible to argue that, in light of WASP-107b’s nearly-polar orbit with respect to the stellar equator, a pathological configuration with $\cos(i_b) \to 0$ is conceivable. In such a scenario, the perturber’s mass, $m_c$ may be assumed to be large enough for von Zeipel–Lidov–Kozai cycles to be activated, allowing $e_b$ to achieve large values. In this regime, the secular Hamiltonian (\ref{Hamiltonian}) becomes a poor approximation to the true dynamics for multiple reasons: (1) the adopted action-angle variables assume small eccentricities and this simplification propagates into both the Hamiltonian and the tidal damping prescription; (2) the assumption that $m_c \ll M_{\star}$ inherent to $\mathcal{H}$ breaks down as the system approaches orthogonality; and (3) secular theory treats semi-major axes as constants, while in reality high eccentricities trigger tidal decay of the orbit.

Instead of extending the secular framework to incorporate these complexities, we directly model the system’s evolution in the high-$i_b$ limit using full N-body simulations. These calculations employ a Bulirsch–Stoer integrator \citep{1992nrfa.book.....P} and include tidal and general relativistic effects following the framework of \citet{2002ApJ...573..829M}. Adopting a tidal quality factor of $Q=30$ (as insinuated by equation \ref{Ltidal}) and using the system parameters enumerated above, we explore the eccentricity evolution of WASP-107b over a range of inclinations from $i_b = 85$ to $89 \deg$.

Figure \ref{fig:ecc}B illustrates the resulting eccentricity time-series. At $i_b = 85\deg$, the orbit remains in a low-eccentricity, limit-cycle state, akin to solutions shown on panel A of the Figure \ref{fig:ecc}. Conversely for $i_b \geqslant 87\deg$, the von Zeipel–Lidov–Kozai mechanism drives $e_b$ to very high values, followed by significant tidal damping and a marked reduction\footnote{E.g., shrinking by more than 40\% in 1 Myr for $i_b = 89\deg$.} in semi-major axis. At $86\deg$, the evolution initially appears more moderate, striking a balance between circularization and extreme excitation. Yet even this solution proves self-limiting: averaged over secular oscillations, mutual inclination declines at a rate of about $0.7\deg$/Myr, destroying the delicate near-orthogonality and invalidating the assumed mass–inclination relation\footnote{Since the RV-derived mass depends on the inclination as $m_c \propto \cos^{-1}(i)$, for near-orthogonal orbits, even a modest reduction in inclination significantly lowers the inferred true mass of planet c, invalidating the scenario in which the mass scale of the perturber was originally dictated by the specific choice of the initial inclination.}. Over longer timescales, the character of the eccentricity evolution resembles that of initially higher-inclination cases (i.e., exhibiting increasingly large eccentricity excursions followed by strong tidal damping and orbital decay).

In other words, while extreme inclinations and high perturber masses can be fine-tuned to produce the required eccentricity, the resulting configuration is intrinsically transient and cannot persist over evolutionary timescales. Overall, this analysis strongly suggests that tidal heating is an unlikely explanation for the inflated radius of WASP-107b, indicating that a distinct mechanism is responsible for maintaining its interior entropy. In some analogy with the hot Jupiter inflation problem discussed in the Introduction, we consider Ohmic dissipation as an alternative explanation.

\section{Ohmic Dissipation}
\label{sec:ohmic}

Fundamentally, the premise behind the Ohmic heating mechanism is that rapid atmospheric circulation within a weakly ionized atmosphere of a planet induces global currents that flow through the entire gaseous envelope, maintaining the interior entropy \citep{BS10,2010ApJ...724..313P,2012ApJ...748L..17H}. The degree of dissipation can be calculated, given a prescription for the large-scale circulation in the atmosphere, $v$, strength of the magnetic field, $B$, as well as the electrical conductivity profile, $\sigma$. While significant uncertainties exist in each of these parameters, to construct our model, we adopt baseline assumptions informed by the published literature, recognizing that the results principally serve as a plausibility estimate rather than a definitive conclusion. We begin by calculating $\sigma$ first.

\subsection{Electrical Conductivity}

The works of \citet{Sing} and \citet{Welbanks} both show that the atmospheric temperature of WASP-107b reaches $\tilde{T}\approx 1550\,$K at a pressure of $\tilde{\mathcal{P}}=1\,$bar -- a condition similar to typical hot Jupiter atmospheres. While this temperature is insufficient to ionize H or He appreciably, alkali metals -- especially K -- can become (partially) thermally ionized. Here, we assume that the enhancement in atmospheric metallicity inferred by \citet{Sing} translates proportionally to the abundance of K, though we note that the atmospheric free electron fraction depends only on the square root of the potassium abundance, somewhat diminishing the sensitivity of the results on this assumption. Additionally, we note that the work of \citet{2021PhRvE.103f3203K} has demonstrated how more sophisticated chemical models can yield higher degrees of ionization than the oft-employed K-only assumption.

\begin{figure}[t]
\centering
\includegraphics[width=\columnwidth]{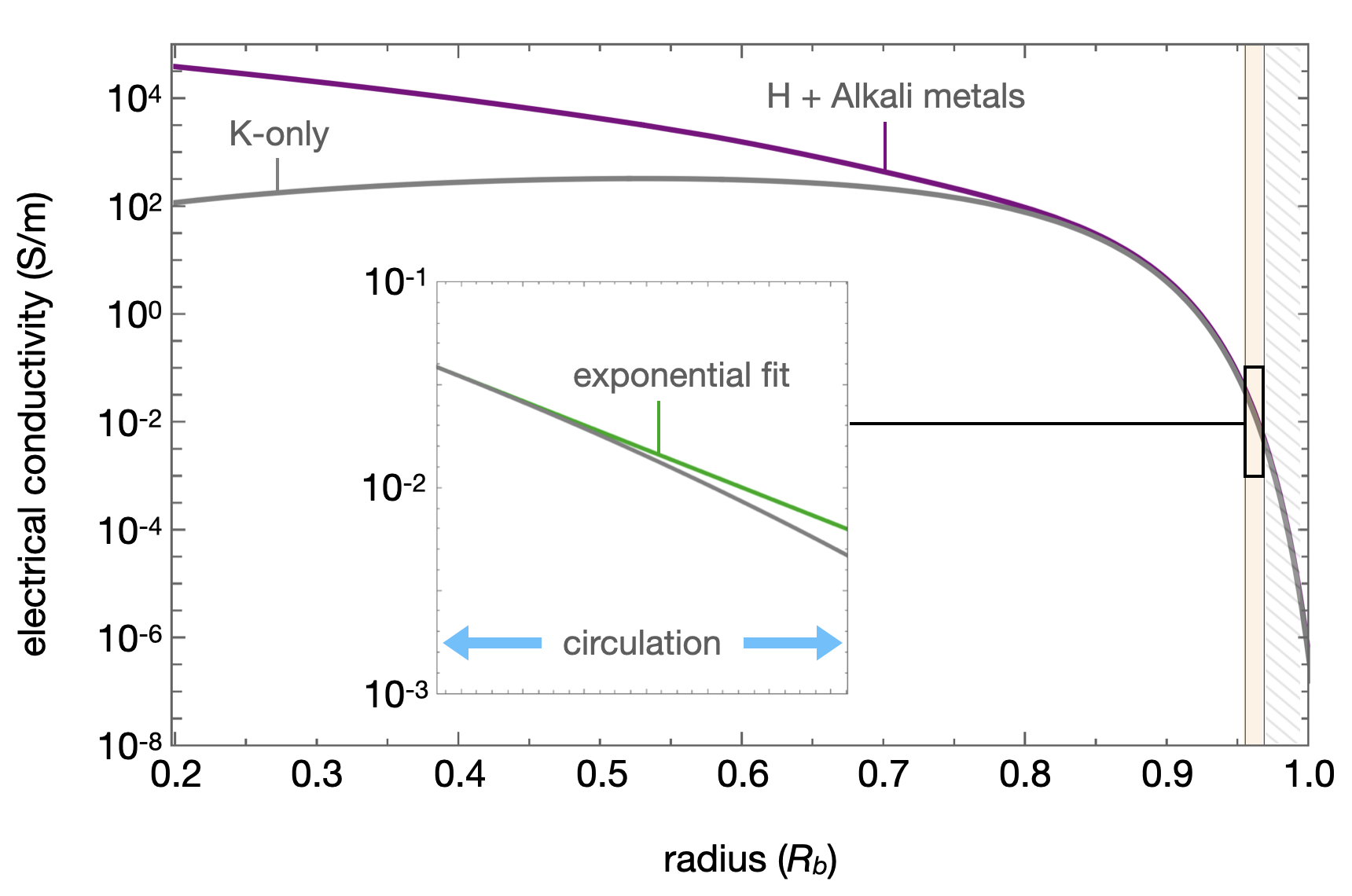}
\caption{Electrical conductivity profile of WASP-107b’s atmosphere and upper interior as a function of radius (in units of $R_b$). The gray curve (labeled ``K-only") shows the conductivity resulting from thermal ionization of potassium alone, while the purple curve (``H + Alkali metals") incorporates multiple alkali species and hydrogen ionization, leading to a significantly higher conductivity in deeper layers. The vertical shaded band on the right marks the portion of the atmosphere with pressure below 1 bar, which we approximate as electrically insulating in the Ohmic heating calculation. The black rectangle denotes the atmospheric circulation zone, where the interactions between the large-scale zonal flow and the ambient magnetic field result in induction. The inset zooms into this region, illustrating an exponential approximation to the conductivity profile (green line), which enables an analytic treatment of the induction equation.}
\label{fig:cond}
\end{figure}

Thermal ionization is governed by the Saha equation, and in the limit of a single ionizing species, the free electron number fraction reads:
\begin{align}
\chi_e\approx \sqrt{\frac{\mu\,m_p\,\chi_K}{\rho}\,\bigg(\frac{m_e\,k_{\rm{b}}\,T}{2\,\pi\,\hbar^2} \bigg)^{3/2}\,\exp\bigg(-\frac{I_K}{k_{\rm{b}}\,T} \bigg)},
\label{Saha}
\end{align}
where $m_p$ and $m_e$ are proton and electron masses respectively, $\mu\approx3.3$ is the mean molecular weight, $\rho$ is the density,  $k_{\rm{b}}$ is the Boltzmann constant, $\hbar$ is the reduced Plank constant, and $I_K=4.34\,$eV is potassium's ionization potential. Moreover, in the above expression, we have assumed a low degree of ionization ($n_{\rm{K}}^{+}\ll n_{\rm{K}} = \chi_K\,n$), with a potassium abundance of $\chi_K \approx 1.4\times 10^{-5}$. 

In the same region (outer envelope) of the planet where the assumption of K-dominated ionization holds, the density profile is closely approximated by an analytic formula (\ref{rhoapprox}). Combining these expressions, we obtain the following equation for the conductivity of the gas:
\begin{align}
\sigma &= \frac{e^2}{4\,\mathcal{A}\,\hbar}\sqrt{\frac{\mu\,m_p\,\chi_K}{\hbar}}\,\bigg(\frac{K\,\mu\,m_p\,m_e}{2\,\pi} \bigg)^{1/4} \nonumber \\
&\ \ \times \bigg(\frac{\eta-1}{\eta}\frac{\G\,m_b}{K\,\tilde{R}}\frac{\tilde{R}-r}{r}+\tilde{\rho}\bigg)^{\frac{\eta-3}{4\,(\eta-1)}} \nonumber \\
&\exp\bigg(-\frac{\eta}{(\tilde{R}-r)(\eta-1)/r+(K/(\G\,m_b))\,\tilde{R}\,\eta\,\tilde{\rho}^{\eta-1}} \nonumber \\
&\ \ \times \frac{I_K\,\tilde{R}}{2\,\G\,m_b\,\mu\,m_p} \bigg),
\label{sigma}
\end{align}
where $\tilde{R}$ and $\tilde{\rho}$ denote the radius and density at the 1 bar pressure-level, which we take as a boundary condition. This equation is depicted with a gray curve as a function of $r$ in Figure \ref{fig:cond}. Though applicable close to the atmosphere, this expression significantly under-estimates the conductivity in deeper region of the planet, where contributions from other metals, as well as Hydrogen itself become non-negligible. Accordingly, as in \citet{BS10,BS11}, we have also computed the full profile of $\sigma$, accounting for the presence of Na, K, Li, Rb, Fe, Cs, and Ca, as well as H (using the tables of \citealt{1995ApJS...99..713S}), and the resulting profile of $\sigma$ is shown as a purple curve in Figure \ref{fig:cond}.



\subsection{Magnetic Field and Atmospheric Circulation}

With the conductivity profile defined, we employ the Ohmic dissipation model outlined in \citet{BS10} to compute the interior heating. Particularly, we adopt a pole-aligned dipolar configuration for the magnetic field, and assume that the large-scale atmospheric circulation manifests as a single zonal jet with a velocity that increases parabolically with altitude:
\begin{align}
&\frac{\vec{B}}{\tilde{B}} = \nabla\times\bigg(\frac{\tilde{R}^3}{r^2}\, \sin(\theta) \,  \hat{\phi}\bigg) \nonumber \\
&\frac{\vec{v}}{\tilde{v}} = \,\bigg(\frac{r-(\tilde{R}-\delta)}{\delta} \bigg)^2\,\sin(\theta) \, \hat{\phi}.
\label{Bv}
\end{align}
where $\delta$ quantifies the depth at which the zonal flow subsides.

To estimate the strength of the magnetic field, we adopt the scaling relation of \citet{2010A&A...522A..13R}, which relates the surface field strength to the heat-flux via the expression:
\begin{align}
\tilde{B} = B_0 \, \bigg( \frac{m_b}{M_{\odot}} \bigg)^{1/6}\, \bigg( \frac{L_b}{L_{\odot}} \bigg)^{1/3}\, \bigg( \frac{R_{\odot}}{R_b} \bigg)^{6/7},
\label{Bsurf}
\end{align}
with a constant of proportionality of $B_0 =0.48\,$T. In essence, this relation is equivalent to the equipartition-like\footnote{That is, it can be derived by setting the kinetic energy density of convective flows equal to the magnetic energy density: $\rho\, v_{\rm{conv}}^2\sim B^2/\mu_0$.} scaling law of \citet{2009Natur.457..167C}  -- appropriate for fully convective, rapidly rotating, spherical dynamos -- where the parameter dependent on the planetary structure is taken to be close to unity. This fiducial estimate yields a surface field of $\tilde{B}\approx70\,$G -- a value comparable to observationally inferred fields of hot Jupiters \citep{2019NatAs...3.1128C}.

Though models of global circulation have not been carried out for WASP-107b specifically, winds in hot and warm Jupiter atmospheres are known to attain peak velocities of a few km/s \citep{2020SSRv..216..139S}. Similarly, simulations of substantially lower equilibrium temperature atmosphere of GJ1214b show the development of $\sim $km/s jets in runs with enhanced ($30\times$ and $50\times$ solar) metallicities \citep{2014ApJ...785...92K}. Furthermore, \citet{2020ApJ...891....7W} have shown that given sufficient GCM convergence time, this jet is expected to extend to a pressure-level of $\sim10\,$bars. 


\begin{figure}[t]
\centering
\includegraphics[width=\columnwidth]{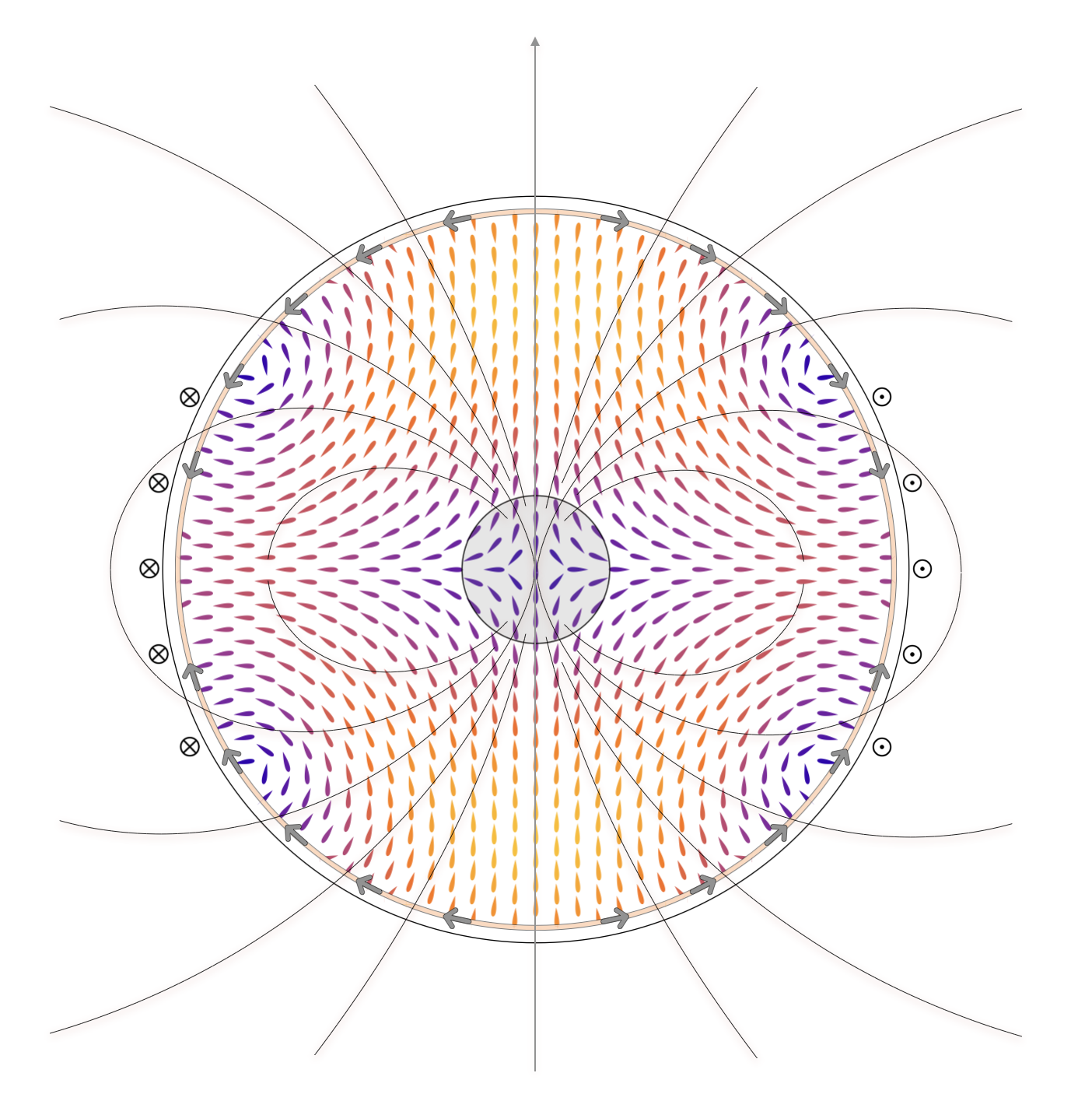}
\caption{Schematic, to-scale representation of the geometry of the induced currents (colored arrows) within WASP-107b’s interior, resulting from atmospheric circulation (indicated by circles with dots and crosses marking flow into and out of the page, respectively) interacting with a pole-aligned magnetic dipole field (black curves). The outer circle denotes the 1-bar pressure level, and the gray band highlights the layer in which zonal winds generate electrical currents. The gray-shaded center represents the planetary core. This global induction pattern facilitates Ohmic heating, maintaining the planet’s internal entropy at an elevated level.}
\label{fig:geom}
\end{figure}

Following these works, we set $\tilde{v}= 0.25\,$km/s at the 1 bar pressure-level, and anchor the bottom\footnote{This specific choice is in part motivated by the atmospheric force balance: between 1 and 2 bar pressure-levels, the Elsasser number, $\Xi = \sigma\,\B^2/(2\,\rho\,\bar{\Omega})$ (where we assume the spin-rate $\bar{\Omega} = n$ to be spin-orbit synchronized), increases from $\Xi\approx0.3$ to $\Xi\approx1.5$, indicating a transition where the Lorentz force becomes dominant over the Coriolis force, thereby diminishing the velocity of the zonal jet \citep{2008Icar..196..653L, 2012ApJ...745..138M, 2014ApJ...782L...4R}} of the circulation layer (where $v\rightarrow0$) at a pressure of 2 bars. At pressures below $\sim1$ bar, temperature and conductivity drop rapidly, so for simplicity, we apply the zero radial current boundary condition at this pressure level. The parameter $\delta$ in equation (\ref{Bv}) thus represents the altitude difference between the 1 and 2 bar pressure levels. We further note that this assumption of a relatively shallow circulation layer participating in induction is a conservative one: extending the wind profile deeper or higher into the atmosphere would result in substantially greater interior heating, as shown by \citet{2022A&A...658L...7K}.

\subsection{Induction and Interior Heating}

The remainder of the calculation is largely identical to that carried out in \citet{BS10} and is explicitly spelled out in the Appendix of that work. Taking advantage of Ohm's law $\vec{J}=\sigma\,(\vec{v}\times \vec{B}-\nabla\,\Phi)$ and the divergence-free nature of the induced current, we obtain the electric potential, $\Phi$, by solving the following equation in the circulation region:
\begin{align}
\nabla \cdot \sigma \, \nabla\,\Phi= \nabla\cdot\vec{v}\times\vec{B},
\label{phicirc}
\end{align}
where we have made the additional assumption that in the circulation region, the conductivity can be approximated by an exponentially decreasing form, matching the conductivity and its radial derivative at the bottom of the weather layer to the polytropic profile (\ref{sigma}) (i.e., $\sigma = \sigma_0 \exp \big( ((r+\delta)-\tilde{R} )(d\sigma/dr)/ \sigma_0 \big)$ -- shown within the inset in Figure {\ref{fig:cond}} as a green curve). This assumption is made because it allows for an analytic solution to the induction equation, thus serving as a convenient parameterization, while providing an acceptable approximation to the polytropic profile across a radially thin layer.


Within the circulation-free region of the interior, the electric potential also follows equation (\ref{phicirc}), but with the RHS set to zero. At $r=0$, we apply the usual Dirichlet boundary condition of $\Phi |_{r=0}=0$, and propagate the solution, assuming that the conductivity of the core material is constant and equal to the value of $\sigma$ at the core-envelope boundary\footnote{Quantitatively, this assumption is unimportant because Ohmic heating is not dominated by the deepest regions of the interior.}. The two solutions are then matched by ensuring continuity of $\Phi$ and $d\Phi/dr$ at the interface (i.e., the 2-bar pressure-level). The geometry of the induced current is depicted in Figure \ref{fig:geom}.

The strength of Ohmic heating is computed by evaluating the volumetric integral of $J^2/\sigma$ within the convective interior (i.e. for $r \leqslant \tilde{R}$). Accordingly, the generated Ohmic power is related to the effective temperature via
\begin{align}
T_{\rm{eff}} = \bigg(\frac{1}{\sigma_{\rm{sb}}\,4\pi\,R_b^2} \int \frac{J^2}{\sigma} d\mathcal{V} \bigg)^{1/4} \propto \sqrt{\tilde{v}\,\tilde{B}}
\label{Teffyo}
\end{align}
With our fiducial estimates enumerated above, we obtain $T_{\rm{eff}}\approx 400\,$K, in agreement with JWST inferences of \citet{Sing,Welbanks}.


\section{Discussion}
\label{sec:discussion}

WASP-107b stands out within the Galactic planetary census as both one of the best-characterized sub-Jovian worlds and one of the least dense. This combination of robust mass, radius, heavy-element content and atmospheric constraints makes it an invaluable laboratory for exploring the physical processes that shape planetary interiors and regulate intrinsic luminosities. A central goal of this work has been to identify a self-consistent mechanism that could power WASP-107b's unexpectedly large interior heat flux and maintain its inflated radius. While previous studies attributed the planet’s high intrinsic heat flux to tidal heating \citep{Mi2020ApJ...897....7M}, our analysis demonstrates that such dissipation is unlikely to account for WASP-107b’s present-day luminosity. Although a non-zero eccentricity is observed, sustaining the high effective temperature of the interior would require a surprisingly low tidal quality factor of $Q_b \sim 30$. Moreover, such vigorous dissipation would act to circularize the orbit on a timescale much shorter than the age of the system, and our modeling of secular dynamics shows that the outer companion cannot maintain the requisite orbital configuration, irrespective of the assumed mutual inclination.

As an alternative, we propose that Ohmic dissipation -- arising from the induction of electrical currents by the interaction of large-scale zonal flows and the planetary magnetic field -- provides a more natural explanation. Previous studies have considered Ohmic dissipation as an energy source in hot Jupiter and mini-Neptune interiors \citep{BS10,2017ApJ...846...47P}, and in this work, we have explored how this mechanism operates under the distinct structural conditions characteristic of super-puff planets. Under fiducial assumptions for the atmospheric circulation speed, magnetic field strength, and ionization chemistry, we find that the resulting Ohmic heating naturally reproduces WASP-107b’s high internal luminosity. Notably, once this alternative heating source is admitted, tides no longer need to supply the observed energy budget, and the bound on $Q_b$ is removed. Thus, interpreting the non-zero eccentricity simply as a relic of slow tidal circularization (on a timescale comparable to the system’s age of a few Gyr; see equation \ref{taue}) pushes the quality factor into the range of $Q_{\rm{b}} \gtrsim \mathrm{few} \times 10^4$ or higher -- consistent with values inferred for Solar System gas and ice giants \citep{2009Natur.459..957L,1988Icar...74..172T}.

We note several assumptions and degeneracies in our modeling. First, our estimate of $k_{2b}$ relies on a simplified interior model consisting of a constant-density core and a polytropic envelope. In principle, a different interior configuration could yield a considerably larger Love number. However, increasing $k_{2b}$ would only weaken the tidal heating argument, since an enhanced quadrupolar precession would further suppress the already insufficient secular eccentricity excitation from the companion, while keeping the circularization timescale (which depends on the ratio $k_{2b}/Q_b$ -- which is determined by the observed luminosity) unchanged. In other words, while our estimates of $k_{2b}$ and $Q_b$ values hinge on structural assumptions, exploring more complex interior models would not qualitatively change our conclusion that tides are inadequate to power WASP-107b’s observed luminosity.

On the Ohmic front, the calculation is deliberately stylized. The total heating rate scales as the square of both the reference-level flow speed and magnetic field strength \citep{BS10}, while also increasing non-trivially with the effective thickness of the circulation layer \citep{2017ApJ...844...94K,2016ApJ...819..116G,2022A&A...658L...7K}. Consequently, one can trade off atmospheric depth, wind speed and profile, as well as magnetic field intensity to achieve effectively identical results. As one example, extending the induction region from 1 to 10 (instead of 2) bars and reducing the peak wind speed from 250 m/s to 40 m/s yields a similar outcome. Likewise, envisioning even deeper circulation\footnote{See the recent work of \citet{2024arXiv240907651W} for an analysis of downward propagation of angular momentum via Lorentz torques.} (e.g., $v\rightarrow0$ at 100 bars) with commensurately slower (only a few m/s at 1 bar) zonal flows also yields an identical integrated heating rate (\ref{Teffyo}). Thus, the combination of parameters provided here should essentially be viewed as a proof-of-concept demonstration, rather than a definitive prescription of WASP-107b’s atmospheric and interior state. That said, if Ohmic dissipation indeed maintains WASP-107b’s inflated radius, it implies that the efficiency of the mechanism in this particular environment is higher ($(T_{\rm{eff}}/T_{\rm{eq}})^4 \lesssim 10\%$) than the typical $\sim 1-3\%$ efficiency inferred for the general hot Jupiter population \citep{BS11,2018AJ....155..214T}. One might speculate that the planet’s high metallicity and extended scale height, as indicated by JWST spectroscopy, promote a more gradual increase in conductivity with depth, yielding an enhanced heating rate.

It is further worth emphasizing that radius inflation in sub-Jovian planets may not have a universal cause. While Ohmic dissipation appears as a compelling explanation for WASP-107b, other super-puff planets could be inflated by alternate processes. Obliquity tides \citep{2019NatAs...3..424M}, enhanced atmospheric opacity that delays contraction \citep{2007ApJ...661..502B}, or other energy sources (\citealt{2002A&A...385..156G,2010ApJ...721.1113Y,2017ApJ...841...30T}; including yet-unidentified ones) may dominate in different systems. Meanwhile, some planets might be inflated simply as a consequence of a low core-to-envelope mass-ratio (see e.g., \citealt{2013ApJ...769L...9B,2014ApJ...792....1L}). As JWST and future missions continue to refine our understanding of planetary atmospheres, compositions, and thermal states, the mechanisms that shape the radii and thermal evolution of planets across the galaxy are bound to come into sharper focus. WASP-107b, serving as a crucial case-study of a benchmark ``super-puff", paves the way for this deeper understanding.


\acknowledgments I am indebted to Daniel Thorngren, whose seminar at Caltech sparked my interest in this system, as well as to Dave Stevenson, Heather Knutson, Andrew Howard, Erik Petigura, Tony Yap, Ian Brunton, Gabriele Pichierri, Matthew Belyakov, Henrik Knierim and Juliette Becker for insightful discussions. I thank the anonymous referee for a thorough and insightful report, which improved the quality of the manuscript. I am grateful to the David and Lucile Packard Foundation, the Caltech Center for Comparative Planetary Evolution (3CPE), and the National Science Foundation (grant number: AST 2408867) for their generous support.

\appendix
\twocolumngrid

\section{Interior Structure}

\paragraph{Hydrostatic Equilibrium} A variety of methods for modeling the interior structures of exoplanets can be found in the literature, each differing in complexity and underlying assumptions (see e.g., \citealt{2013ApJ...769L...9B,2016ApJ...825...29G,2016ApJ...831..180C,2024ApJ...977..227K} and the references therein). Here, we adopt a simplified, but time-honored approach of using a polytropic equation of state \citep{1939isss.book.....C}. Specifically, for the gaseous envelope, we assume the pressure-density relation:
\begin{align} 
\mathcal{P} = K\, \rho^{\eta},
\label{polytrope}
\end{align} 
where the adiabatic index is related to the polytropic index via $\eta = 1+1/\zeta$. At the $\tilde{\mathcal{P}} = 1\,$bar pressure level -- taken as the nominal radiative-convective boundary in the detailed models of \citet{Sing,Welbanks} -- WASP-107b’s temperature is inferred to be $\tilde{T}\approx1550\,\mathrm{K}$. From the ideal gas law, we thus have: $\tilde{\rho} = \tilde{\mathcal{P}}\,\mu\,m_p/(k_{\rm{b}}\,\tilde{T})$, and the value of the constant in \ref{polytrope} follows from: $K = \tilde{\mathcal{P}}/\tilde{\rho}^{\eta}$. Given the $\sim 43\times$ solar metallicity of the envelope derived by \citet{Sing}, we adopt a mean-molecular weight\footnote{For comparison, a solar composition H/He mixture has a mean molecular weight of $\mu \approx 2.4$, while the Earth's atmosphere has $\mu \approx 29$.} of $\mu = 3.3$, in agreement with previous work \citep{2013ApJ...775...80F}. 

Generally, the structure of the planet is governed by the equation of hydrostatic equilibrium,
\begin{align}
\frac{d\mathcal{P}}{dr} = -\rho\, g, 
\label{hydrostatic1}
\end{align} where $g$ is the local gravitational acceleration. Combining this expression with the polytropic relation (\ref{polytrope}), and taking a derivative with respect to $r$, yields a second-order ordinary differential equation:
\begin{align}
(\eta-2)\,\frac{r}{\rho}\,\bigg(\frac{d\rho}{dr} \bigg)^2+2\,\frac{d\rho}{dr}+r\,\frac{d^2\rho}{dr^2} = \frac{4\,\pi\,\G}{K\,\eta}\frac{r}{\rho^{\eta-3}}.
\label{hydrostatic2}
\end{align}
To solve this ODE, two boundary conditions must be specified. At the core-envelope interface ($r=R_{\rm{c}}$), the Neumann boundary condition on the density gradient follows directly from equation (\ref{hydrostatic1}): $d\rho/dr |_{R_{\rm{c}}}= -\G\,M_{\rm{c}}\,\rho_{0}^{2-\eta}/(K\,\eta\,R_{\rm{c}}^2)$. Additionally, a Dirichlet boundary condition sets the (maximum) density of the polytropic envelope at the same location: $\rho|_{R_{\rm{c}}}=\rho_{0}$. The core radius is determined by adopting an Earth-like composition and a Murnaghan EOS for the core material, which implies a simple mass-radius relation: $R_{\rm{c}}/R_{\oplus} = (M_{\rm{c}}/M_{\oplus})^{1/4}$, where we set $M_{\rm{c}}$ to $12 M_{\oplus}$.

\begin{figure}
\includegraphics[width=\columnwidth]{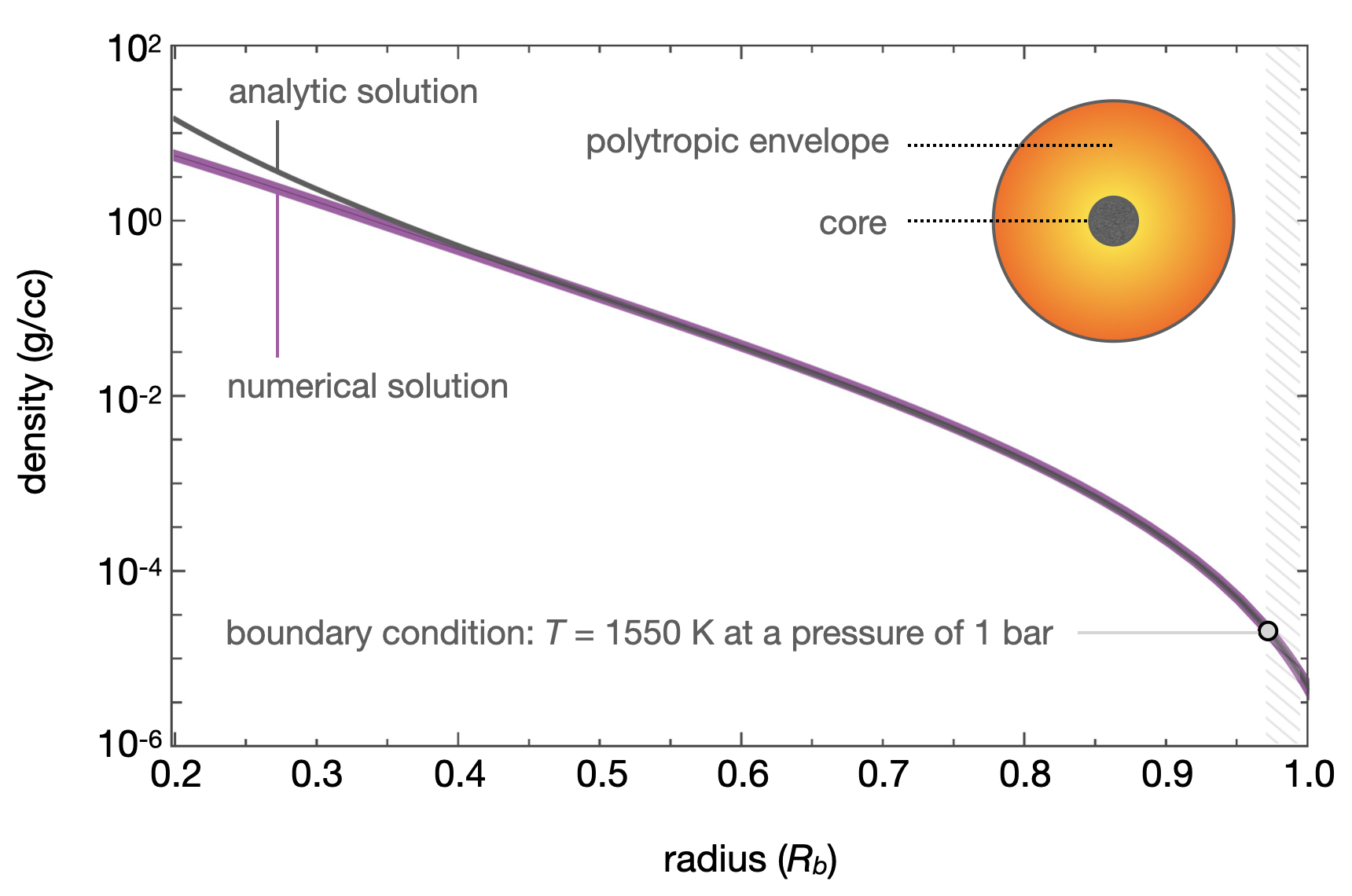}
\caption{Density profile of WASP-107b’s polytropic envelope, extending from the core to the outer radius, defined by matching the atmospheric temperature the irradiation equilibrium value. The purple curve shows the full numerical hydrostatic solution, while the black curve represents the analytic approximation given by equation (\ref{rhoapprox}). Both solutions assume a boundary condition corresponding to a temperature of $\tilde{T}=1550\,$K at the 1-bar pressure level (denoted by an open circle). Due to the planet’s highly centrally concentrated interior (quantified by a Love number $k_{2b}=0.015$), the analytic model offers an excellent approximation to the numerical solution throughout most of the interior.}
\label{fig:interior}
\end{figure}

Within the polytropic envelope, the temperature is related to density via $T = K\,\mu\,m_p\,\rho^{\eta - 1}/k_{\rm{b}}$. We define the planetary radius $R_b$ as the location where the envelope’s temperature drops to the equilibrium irradiation temperature\footnote{We note that this approximation neglects the markedly non-polytropic structure of the planetary atmosphere. However, this has little practical consequence for the determination of the radius.}, $T_{\rm{irr}}=770\,\mathrm{K}$. For a given value of $\eta$, adjusting $\rho_0$ yields a mass-radius relation, and matching WASP-107b’s measured total mass and radius uniquely pinpoints the values of both variables to $\eta=1.29$ and $\rho_0=5.8\,$g/cc. We remark that this value of $\eta$ implies a polytropic index of $\zeta\approx 7/2$ and the value of $\rho_0$ is comparable to the inferred density at the core-envelope interface of Uranus and Neptune (\citealt{Militzer}; see also \citealt{2000P&SS...48..143P}). The solution is shown as purple curve in Figure (\ref{fig:interior}).

\paragraph{Love Number} With the planet’s internal density distribution in hand, we quantify its degree of central concentration through the dimensionless Love number\footnote{Physically, $k_{2b}$ characterizes the planet’s response to tidal deformation.} $k_{2b}$. Following \citet{1939MNRAS..99..451S}, the computation of $k_{2b}$ begins by defining the mean density interior to a radius $r$: 
\begin{align}
\bar{\rho} = \frac{3}{4\pi\,r^3} \bigg(M_{\rm{c}} + \int_{R_{\rm{c}}}^r 4\pi\,\rho\,r'^2\,dr' \bigg).
\label{rhobar}
\end{align} 
Then, we integrate the following second-order ordinary differential equation for the variable $\varsigma_2$ outward from the core radius:
\begin{align} 
r\,\frac{d\varsigma_2}{dr}+\varsigma_2^2-\varsigma_2-6+\frac{6\,\rho}{\bar{\rho}}\big(\varsigma_2+1 \big)=0,
\end{align} 
with the boundary condition\footnote{This assumes a constant density core. For a variable density core, the procedure is the same, but the integration should be carried out from the center of the planet, with $M_{\rm{c}} $ dropped from the integral in (\ref{rhobar}).} $\varsigma_2|_{R_{\rm{c}}}=0$. Once integrated to the planetary radius $R_b$, the Love number is obtained from:
\begin{align} 
k_{2b} = \frac{3-\varsigma_2|_{R_b}}{2+\varsigma_2|_{R_b}}.
\end{align} 
The polytropic model of the interior delineated above yields $k_{2b}=0.015$.

\paragraph{Approximate Solution} Because the mass distribution is strongly concentrated towards the core, the majority of the envelope experiences a nearly constant enclosed mass, and thus the local gravitational acceleration can be approximated by $g \approx \G m_b/r^2$. Under this approximation, the hydrostatic equation reduces to a simpler, first-order differential equation, allowing for a fully analytic solution once the density at the radiative-convective boundary, $\tilde{\rho}$, is specified:
\begin{align}
\rho \approx \Bigg(\tilde{\rho}^{\eta-1}-\frac{\eta-1}{\eta} \frac{\G\,m_b\,}{K\,\tilde{R}}\frac{\big(r-\tilde{R} \big)}{r} \Bigg)^{\frac{1}{\eta-1}}.
\label{rhoapprox}
\end{align}
This analytic solution, shown as the black curve in Figure \ref{fig:interior}, closely tracks the full numerical solution throughout the bulk of the envelope.

\section{Secular Perturbation Theory}

The orbital architecture of the WASP-107 system is strongly hierarchical, and no meaningful mean-motion commensurabilities are evident. Consequently, the long-term evolution of the planetary orbits can be captured using a secular (orbit-averaged) approximation to the dynamics. To this end, we adopt a Kaula-like expansion of the disturbing function. Truncated at octupole order, and expressed in terms of Keplerian orbital elements, the resulting secular Hamiltonian reads \citep{2010MNRAS.407.1048M}:
 \begin{align}
&\mathcal{H}'=-\frac{1}{4}\frac{\G\,m_b\,m_c}{a_b}\bigg(\frac{a_b}{a_c\,\epsilon_c} \bigg)^3 \Bigg[ \bigg(1+\frac{3\,e_b^2}{2} \bigg) \nonumber \\
&\times \frac{3\,\cos^2(i_b)-1}{2}\frac{3\,\cos^2(i_c)-1}{2}+\frac{15}{4}\,e_b^2\,\sin^2(i_b)\,\cos(2 \omega) \nonumber \\
&+\frac{3}{4}\,\sin(2i_b)\,\sin(2i_c)\,\cos(\Omega_b-\Omega_c) \nonumber \\ 
&+\frac{3}{4}\,\sin^2(i_b)\,\sin^2(i_c)\,\cos(2\Omega_b-2\Omega_c)\Bigg] \nonumber \\ 
&+\frac{15}{16}\frac{\G\,m_b\,m_c}{a_b} \frac{e_b\,e_c}{\epsilon_c} \bigg(\frac{a_b}{a_c\,\epsilon_c} \bigg)^4 \Bigg[ \cos(\varpi_b-\varpi_c) \nonumber \\ 
&\times\frac{1+\cos(i_b)}{8}\,(15\,\cos^2(i_b)-10\,\cos(i_b) -1) \nonumber \\ 
&+\frac{1-\cos(i_b)}{8}\,(15\,\cos^2(i_b)+10\,\cos(i_b) -1) \nonumber \\ 
&\times\cos(\varpi_b+\varpi_2-2\Omega_b) \Bigg]
 \label{Hfull}
 \end{align}
Notably, this flavor of secular perturbation theory places no restrictions on eccentricities or inclinations but does assume that $m_b, m_c \ll M_{\star}$ and that $\alpha = a_b/a_c \ll 1$.

Although the true mass of planet c is unknown, even its minimum mass ensures that the ratio of angular momenta satisfies 
\begin{align}
\frac{m_b}{m_c}\,\sqrt{\frac{a_b\,(1-e_b^2)}{a_c\,(1-e_c^2)}} \ll 1.
\end{align} 
Under this condition, perturbations exerted onto the outer orbit by the inner orbit are effectively negligible. We therefore choose a reference plane aligned with planet c's orbit and consider $e_c$ and $i_c$ as nearly constant. This eliminates terms proportional to $\sin(i_c)$ and removes $\Omega_c$ from the problem entirely. As such, the dynamical evolution of the outer orbit is reduced to slow apsidal precession, and the associated rate can be readily computed from Lagrange's planetary equations applied to (\ref{Hfull}), yielding (to quadrupole order) a simple expression given in equation (\ref{varpic}).

To facilitate a Hamiltonian formulation, we employ canonically conjugated Poincar\'e action-angle variables:
\begin{align}
&\Lambda'=m_b\,\sqrt{\G\,M_{\star}\,a_b} &\lambda = \mathcal{M}_b+\varpi_b \nonumber \\
&\Gamma'=\Lambda\,\big(1-\sqrt{1-e_b^2} \big) &\gamma = -\varpi_b \nonumber \\
&Z'=\big(\Lambda-\Gamma\big)\big(1-\cos(i_b) \big) &z = -\Omega_b
\end{align} 
Because semi-major axes remain invariant in secular theory, $\Lambda'$ is a constant parameter rather than a dynamic variable. By rescaling the actions and the Hamiltonian with $\Lambda'$ (such that $\mathcal{H}=\mathcal{H}'/\Lambda'$), and expanding the definitions of the actions to leading order in $e_b$, we obtain a simplified set of action-angle variables that describe the inner orbit’s eccentricity and spatial orientation:
\begin{align}
&\Gamma\approx e_b^2/2 &Z\approx \big(1-\cos(i_b) \big).
\end{align} 
Expressing the scaled Hamiltonian in terms of these actions, and adding leading-order corrections for relativistic and tidal contributions, produces the final form of the secular model as presented in the main text (equation \ref{Hamiltonian}). 



\begin{thebibliography}


\bibitem[Arras \& Socrates(2010)]{2010ApJ...714....1A} Arras, P. \& Socrates, A.\ 2010, \apj, 714, 1. doi:10.1088/0004-637X/714/1/1



\bibitem[Batygin et al.(2009)]{2009ApJ...704L..49B} Batygin, K., Bodenheimer, P., \& Laughlin, G.\ 2009, \apjl, 704, L49. doi:10.1088/0004-637X/704/1/L49

\bibitem[Batygin \& Stevenson(2010)]{BS10} Batygin, K. \& Stevenson, D.~J.\ 2010, \apjl, 714, L238. doi:10.1088/2041-8205/714/2/L238

\bibitem[Batygin et al.(2011)]{BS11} Batygin, K., Stevenson, D.~J., \& Bodenheimer, P.~H.\ 2011, \apj, 738, 1. doi:10.1088/0004-637X/738/1/1

\bibitem[Batygin \& Stevenson(2013)]{2013ApJ...769L...9B} Batygin, K. \& Stevenson, D.~J.\ 2013, \apjl, 769, L9. doi:10.1088/2041-8205/769/1/L9


\bibitem[Bodenheimer et al.(2001)]{2001ApJ...548..466B} Bodenheimer, P., Lin, D.~N.~C., \& Mardling, R.~A.\ 2001, \apj, 548, 466. doi:10.1086/318667

\bibitem[Burrows et al.(2007)]{2007ApJ...661..502B} Burrows, A., Hubeny, I., Budaj, J., et al.\ 2007, \apj, 661, 502. doi:10.1086/514326



\bibitem[Cauley et al.(2019)]{2019NatAs...3.1128C} Cauley, P.~W., Shkolnik, E.~L., Llama, J., et al.\ 2019, Nature Astronomy, 3, 1128. doi:10.1038/s41550-019-0840-x

\bibitem[Charbonneau et al.(2000)]{2000ApJ...529L..45C} Charbonneau, D., Brown, T.~M., Latham, D.~W., et al.\ 2000, \apjl, 529, L45. doi:10.1086/312457

\bibitem[Chandrasekhar(1939)]{1939isss.book.....C} Chandrasekhar, S.\ 1939, Chicago, Ill., The University of Chicago press [1939]


\bibitem[Chen \& Rogers(2016)]{2016ApJ...831..180C} Chen, H. \& Rogers, L.~A.\ 2016, \apj, 831, 180. doi:10.3847/0004-637X/831/2/180


\bibitem[Christensen et al.(2009)]{2009Natur.457..167C} Christensen, U.~R., Holzwarth, V., \& Reiners, A.\ 2009, \nat, 457, 167. doi:10.1038/nature07626




\bibitem[Dai \& Winn(2017)]{2017AJ....153..205D} Dai, F. \& Winn, J.~N.\ 2017, \aj, 153, 205. doi:10.3847/1538-3881/aa65d1




\bibitem[Fortney et al.(2013)]{2013ApJ...775...80F} Fortney, J.~J., Mordasini, C., Nettelmann, N., et al.\ 2013, \apj, 775, 80. doi:10.1088/0004-637X/775/1/80


\bibitem[Fuller et al.(2016)]{2016MNRAS.458.3867F} Fuller, J., Luan, J., \& Quataert, E.\ 2016, \mnras, 458, 3867. doi:10.1093/mnras/stw609



\bibitem[Ginzburg et al.(2016)]{2016ApJ...825...29G} Ginzburg, S., Schlichting, H.~E., \& Sari, R.\ 2016, \apj, 825, 29. doi:10.3847/0004-637X/825/1/29

\bibitem[Ginzburg \& Sari(2016)]{2016ApJ...819..116G} Ginzburg, S. \& Sari, R.\ 2016, \apj, 819, 116. doi:10.3847/0004-637X/819/2/116


\bibitem[Goldberg \& Batygin(2024)]{2024Icar..41316014G} Goldberg, M. \& Batygin, K.\ 2024, \icarus, 413, 116014. doi:10.1016/j.icarus.2024.116014


\bibitem[Guillot \& Showman(2002)]{2002A&A...385..156G} Guillot, T. \& Showman, A.~P.\ 2002, \aap, 385, 156. doi:10.1051/0004-6361:20011624



\bibitem[Heng(2012)]{2012ApJ...748L..17H} Heng, K.\ 2012, \apjl, 748, L17. doi:10.1088/2041-8205/748/1/L17


\bibitem[Henry et al.(2000)]{2000ApJ...529L..41H} Henry, G.~W., Marcy, G.~W., Butler, R.~P., et al.\ 2000, \apjl, 529, L41. doi:10.1086/312458





\bibitem[Kataria et al.(2014)]{2014ApJ...785...92K} Kataria, T., Showman, A.~P., Fortney, J.~J., et al.\ 2014, \apj, 785, 92. doi:10.1088/0004-637X/785/2/92

\bibitem[Knierim et al.(2022)]{2022A&A...658L...7K} Knierim, H., Batygin, K., \& Bitsch, B.\ 2022, \aap, 658, L7. doi:10.1051/0004-6361/202142588

\bibitem[Knierim \& Helled(2024)]{2024ApJ...977..227K} Knierim, H. \& Helled, R.\ 2024, \apj, 977, 227. doi:10.3847/1538-4357/ad8dd0

\bibitem[Komacek \& Youdin(2017)]{2017ApJ...844...94K} Komacek, T.~D. \& Youdin, A.~N.\ 2017, \apj, 844, 94. doi:10.3847/1538-4357/aa7b75


\bibitem[Kumar et al.(2021)]{2021PhRvE.103f3203K} Kumar, S., Poser, A.~J., Sch{\"o}ttler, M., et al.\ 2021, \pre, 103, 063203. doi:10.1103/PhysRevE.103.063203



\bibitem[Laughlin et al.(2011)]{2011ApJ...729L...7L} Laughlin, G., Crismani, M., \& Adams, F.~C.\ 2011, \apjl, 729, L7. doi:10.1088/2041-8205/729/1/L7

\bibitem[Lainey et al.(2009)]{2009Natur.459..957L} Lainey, V., Arlot, J.-E., Karatekin, {\"O}., et al.\ 2009, \nat, 459, 957. doi:10.1038/nature08108

\bibitem[Lainey et al.(2020)]{2020NatAs...4.1053L} Lainey, V., Casajus, L.~G., Fuller, J., et al.\ 2020, Nature Astronomy, 4, 1053. doi:10.1038/s41550-020-1120-5

\bibitem[Liu et al.(2008)]{2008Icar..196..653L} Liu, J., Goldreich, P.~M., \& Stevenson, D.~J.\ 2008, \icarus, 196, 653. doi:10.1016/j.icarus.2007.11.036

\bibitem[Lopez \& Fortney(2014)]{2014ApJ...792....1L} Lopez, E.~D. \& Fortney, J.~J.\ 2014, \apj, 792, 1. doi:10.1088/0004-637X/792/1/1



\bibitem[Mardling \& Lin(2002)]{2002ApJ...573..829M} Mardling, R.~A. \& Lin, D.~N.~C.\ 2002, \apj, 573, 829. doi:10.1086/340752


\bibitem[Mardling(2007)]{2007MNRAS.382.1768M} Mardling, R.~A.\ 2007, \mnras, 382, 1768. doi:10.1111/j.1365-2966.2007.12500.x

\bibitem[Mardling(2010)]{2010MNRAS.407.1048M} Mardling, R.~A.\ 2010, \mnras, 407, 1048. doi:10.1111/j.1365-2966.2010.16814.x

\bibitem[Mayor \& Queloz(1995)]{1995Natur.378..355M} Mayor, M. \& Queloz, D.\ 1995, \nat, 378, 355. doi:10.1038/378355a0

\bibitem[Menou(2012)]{2012ApJ...745..138M} Menou, K.\ 2012, \apj, 745, 138. doi:10.1088/0004-637X/745/2/138

\bibitem[Militzer(2024)]{Militzer} Militzer, B. \ 2024, PNAS, 121, 49 doi:10.1073/pnas.2403981121

\bibitem[Millholland \& Laughlin(2019)]{2019NatAs...3..424M} Millholland, S. \& Laughlin, G.\ 2019, Nature Astronomy, 3, 424. doi:10.1038/s41550-019-0701-7

\bibitem[Millholland et al.(2020)]{Mi2020ApJ...897....7M} Millholland, S., Petigura, E., \& Batygin, K.\ 2020, \apj, 897, 7. doi:10.3847/1538-4357/ab959c





\bibitem[Perna et al.(2010)]{2010ApJ...724..313P} Perna, R., Menou, K., \& Rauscher, E.\ 2010, \apj, 724, 313. doi:10.1088/0004-637X/724/1/313

\bibitem[Petigura et al.(2020)]{2020AJ....159....2P} Petigura, E.~A., Livingston, J., Batygin, K., et al.\ 2020, \aj, 159, 2. doi:10.3847/1538-3881/ab5220

\bibitem[Piaulet et al.(2021)]{2021AJ....161...70P} Piaulet, C., Benneke, B., Rubenzahl, R.~A., et al.\ 2021, \aj, 161, 70. doi:10.3847/1538-3881/abcd3c

\bibitem[Press et al.(1992)]{1992nrfa.book.....P} Press, W.~H., Teukolsky, S.~A., Vetterling, W.~T., et al.\ 1992, Cambridge: University Press, |c1992, 2nd ed.

\bibitem[Podolak et al.(2000)]{2000P&SS...48..143P} Podolak, M., Podolak, J.~I., \& Marley, M.~S.\ 2000, \planss, 48, 143. doi:10.1016/S0032-0633(99)00088-4

\bibitem[Pollack et al.(1996)]{1996Icar..124...62P} Pollack, J.~B., Hubickyj, O., Bodenheimer, P., et al.\ 1996, \icarus, 124, 62. doi:10.1006/icar.1996.0190

\bibitem[Pu \& Valencia(2017)]{2017ApJ...846...47P} Pu, B. \& Valencia, D.\ 2017, \apj, 846, 47. doi:10.3847/1538-4357/aa826f




\bibitem[Ragozzine \& Wolf(2009)]{2009ApJ...698.1778R} Ragozzine, D. \& Wolf, A.~S.\ 2009, \apj, 698, 1778. doi:10.1088/0004-637X/698/2/1778

\bibitem[Reiners \& Christensen(2010)]{2010A&A...522A..13R} Reiners, A. \& Christensen, U.~R.\ 2010, \aap, 522, A13. doi:10.1051/0004-6361/201014251

\bibitem[Rogers \& Showman(2014)]{2014ApJ...782L...4R} Rogers, T.~M. \& Showman, A.~P.\ 2014, \apjl, 782, L4. doi:10.1088/2041-8205/782/1/L4


\bibitem[Rubenzahl et al.(2021)]{2021AJ....161..119R} Rubenzahl, R.~A., Dai, F., Howard, A.~W., et al.\ 2021, \aj, 161, 119. doi:10.3847/1538-3881/abd177



\bibitem[Sarkis et al.(2021)]{2021A&A...645A..79S} Sarkis, P., Mordasini, C., Henning, T., et al.\ 2021, \aap, 645, A79. doi:10.1051/0004-6361/202038361

\bibitem[Saumon et al.(1995)]{1995ApJS...99..713S} Saumon, D., Chabrier, G., \& van Horn, H.~M.\ 1995, \apjs, 99, 713. doi:10.1086/192204

\bibitem[Showman et al.(2020)]{2020SSRv..216..139S} Showman, A.~P., Tan, X., \& Parmentier, V.\ 2020, \ssr, 216, 139. doi:10.1007/s11214-020-00758-8

\bibitem[Sing et al.(2024)]{Sing} Sing, D.~K., Rustamkulov, Z., Thorngren, D.~P., et al.\ 2024, \nat, 630, 831. doi:10.1038/s41586-024-07395-z

\bibitem[Sterne(1939)]{1939MNRAS..99..451S} Sterne, T.~E.\ 1939, \mnras, 99, 451. doi:10.1093/mnras/99.5.451



\bibitem[Thorngren \& Fortney(2018)]{2018AJ....155..214T} Thorngren, D.~P. \& Fortney, J.~J.\ 2018, \aj, 155, 214. doi:10.3847/1538-3881/aaba13

\bibitem[Tittemore \& Wisdom(1988)]{1988Icar...74..172T} Tittemore, W.~C. \& Wisdom, J.\ 1988, \icarus, 74, 172. doi:10.1016/0019-1035(88)90038-3



\bibitem[Tremblin et al.(2017)]{2017ApJ...841...30T} Tremblin, P., Chabrier, G., Mayne, N.~J., et al.\ 2017, \apj, 841, 30. doi:10.3847/1538-4357/aa6e57





\bibitem[Wang \& Wordsworth(2020)]{2020ApJ...891....7W} Wang, H. \& Wordsworth, R.\ 2020, \apj, 891, 7. doi:10.3847/1538-4357/ab6dcc

\bibitem[Wazny \& Menou(2024)]{2024arXiv240907651W} Wazny, M. \& Menou, K.\ 2024, arXiv:2409.07651. doi:10.48550/arXiv.2409.07651

\bibitem[Welbanks et al.(2024)]{Welbanks} Welbanks, L., Bell, T.~J., Beatty, T.~G., et al.\ 2024, \nat, 630, 836. doi:10.1038/s41586-024-07514-w


\bibitem[Wolszczan \& Frail(1992)]{1992Natur.355..145W} Wolszczan, A. \& Frail, D.~A.\ 1992, \nat, 355, 145. doi:10.1038/355145a0




\bibitem[Youdin \& Mitchell(2010)]{2010ApJ...721.1113Y} Youdin, A.~N. \& Mitchell, J.~L.\ 2010, \apj, 721, 1113. doi:10.1088/0004-637X/721/2/1113

\bibitem[Yu \& Dai(2024)]{2024ApJ...972..159Y} Yu, H. \& Dai, F.\ 2024, \apj, 972, 159. doi:10.3847/1538-4357/ad5ffb




\end{thebibliography}

\bibliographystyle{aasjournal}

\end{document}